# A Density Functional Theory Study of Magnetic Transition in MnO$_2$ adsorbed Vanadium Carbide (V$_2$C) MXene

Mahjabeen Fatima[1,2], Saleem Ayaz Khan[3*], Syed Rizwan[1*]

[1]Physics Characterization and Simulations Lab (PCSL), Department of Physics, School of Natural Sciences (SNS), National University of Sciences and Technology (NUST), Islamabad 44000, Pakistan.

[2]Department of Physics and Astronomy, School of Natural Sciences, The University of Manchester, United Kingdom.

[3]New Technologies Research Centre, University of West Bohemia, Univerzitni 2732, 306 14 Pilsen, Czech Republic.

**Corresponding authors:** 1) Syed Rizwan: syedrizwan@sns.nust.edu.pk, syedrizwanh83@gmail.com, 2) Saleem Ayaz Khan: sayaz_usb@yahoo.com

**Abstract**

The work reports nonmagnetic behavior (0.04 µB) in two-dimensional (2D) V$_2$C-OF MXene and ferromagnetism in MnO$_2$ adsorbed V$_2$C-OF MXene. The density functional theory (DFT) calculations were carried out to study the magnetic moments of V$_2$C-OF and MnO$_2$@V$_2$C-OF MXene. The MXene, which is derived from the exfoliation of its parent V$_2$AlC MAX phase, shows a good potential to be a ferromagnetic when MnO$_2$ is adsorbed on it. The V$_2$C MXene and MnO$_2$ adsorbed V$_2$C MXene were successfully synthesized, as characterized using X-ray diffraction, showing an increased c-lattice parameter from 22.6Å to 27.2Å after MnO$_2$ adsorption. The DFT study confirmed that MnO$_2$ adsorbed V$_2$C MXene changed from nonmagnetic (in V$_2$C MXene) to a strong ferromagnetic with a magnetic moment of 4.48$\mu_B$ for Mn adsorbed V$_2$C-OF MXene. The current work is a step-forward towards understanding of magnetism in two-dimensional materials for future 2D spintronics.

## Introduction

The tuning of electronic and magnetic properties of a material through adsorption of one element or compound over varying class of MXenes is an effective strategy for the enhancement in energy storage systems and spintronic devices [1–3]. With the growing requirement of novel and smart materials, several compounds have been engineered for the development of nanotechnology industry. The first 2D material, graphene [4], showed intriguing properties for diverse applications after which, many other 2D materials like borophene, hexagonal boron-nitride, phosphorene, transition metal dichalcogenide and phosphorene bismuthene were discovered and are used in the applications of biosensors, hydrogen evolution reactions (HER), photonics, energy storage systems, etc [5–10].

The 2D MXenes are a result of exfoliated MAX phase which is a bulk, three-dimensional material that represents a huge (60+ members) family of transition metal carbides, nitrides, and carbonitrides. These possess lamellar hexagonally symmetric structures (space group P63/mmc) with the generalized formula of $M_{n+1}AX_n$ (n = 1,2,3), whereas 'M' represents a primary transition metal (Ti, Nb, Ta, Mo, V and many more), 'A' shows element from groups III-A and IV-A of the periodic table, and 'X' symbolizes carbon or nitrogen [11–15]. In order to get the MXene, the chemical etching of MAX phase is carried out using a suitable chemical etchant to remove the A-layer from MAX [16,17]. The examples of MXene include $V_2C$, $Nb_2C$, $Cr_2C$, $Ti_2C$, $Ti_3C_2$, $Nb_4C_3$, $Hf_2C$, $Mo_2C$, etc [18, 19]. MXenes involve $(n + 1)$ M layers that enfolds 'n' layers of X in an $[MX]_nM$ sequence. Moreover, chemically-etched pristine MXenes are hard to exist because of its highly reactive surface which results in absorbance of moisture from the atmosphere and may form a bonding with oxygen, hydroxide, fluorine (from chemical etchant), or an oxyfluoride (-OH, -O,-F, -OF) and are typically named as surface terminations ($T_x$) as they tend to lower the reactivity of the MXenes [20,21].

Owing to the presence of transition metals which possess valence electrons in d-orbitals and tendency to form the bonds, MXenes become very important material candidate for their magnetic properties. Non-magnetism, ferromagnetism and anti-ferromagnetism of various kinds have been studied and predicted in MXene families which changes with presence of M element [22]. Gao et al. discussed monolayer $Ti_2C$ and $Ti_2N$, through first-principles calculations, that exhibited half-metallic ferromagnetism. They also discussed that $V_2C$ and $V_2N$ exhibit a non-magnetic nature [23]. Khazaei et al. calculated the magnetic moment (M) of $Cr_2C$ and $Cr_2N$ which showed a narrow band gap of semiconducting nature revealing its ferromagnetism [24]. Shien et al. elaborated the structure of $Ti_{n+1}AlC_n$ and $Ti_{n+1}C_n$ MXenes along with the formation energy required for $Ti_2C$ and $Ti_3C_2$ [25]. Their results indicated increased stability with a higher Ti–C bond count, with Al-containing nano-blocks proving more stable and exhibiting possible magnetic ordering in the Ti layers.

Enyashin et al. studied the effect of functional groups on magnetism of different MXenes using LDA+U, GGA+U and PBEsol functional and reported that the presence of functional groups modifies the spin-orbit coupling of 5d-orbital transition metal that results in the presence of ferrimagnetism and anti-ferromagnetism in $Ta_3C_2$, anti-FM in $Ta_3C_2$, FM and anti-FM in $Ta_2C$ [26, 27]. Zhao et al. discussed the magnetism in different compounds of $M_2C$ MXene under various mechanical strains showing the Meissner effect [28]. Recently, the Nb-doped as well as lanthanides-doped $Ti_3C_2$ MXene showed variable ferromagnetism and anti-ferromagnetism [29–31]. Herein, we used first-principles calculation for computational analysis of magnetic properties of well-prepared $V_2C$ and $MnO_2$ adsorbed $V_2C$ MXene using density functional theory (DFT) and discussed the effect on their magnetic properties that changes from non-magnetic (in $V_2C$) to ferromagnetic (in $MnO_2$-adsorbed $V_2C$ MXene).

**Experimental Details**

In Figure 1, the schematic of selective etching of Al layer from $V_2AlC$ to obtain $V_2CT_x$ is shown. The synthesis of $V_2CT_x$ MXene was initiated by taking 1g of $V_2AlC$ MAX (300 mesh) and treating it with the chemical etchant (49% concentrated hydrofluoric acid, ACS grade, BDH) for optimized 116 hours at room-temperature. Magnetic stirring was constantly carried out by a Teflon-coated magnetic stirrer at 300 rpm. Etched sample was washed 4 to 5 times by using DI water and ethanol after centrifugation at 4500 rpm till the supernatant obtained a pH of 5. MXene was then filtered out with the help of vacuum filtration process in which, the solution was rinsed through DI water along with absolute ethanol using a celgard porous membrane having a pore size of 0.22μm. Powder sample of $V_2C$ MXene was obtained after drying in a vacuum oven for 24 hours.

$MnO_2$-$V_2C$ nanocomposite was synthesized by liquid-phase precipitation method at high temperature. At first, 200 mg of $V_2C$ powder was dispersed in 100 mL, 1 mM aqueous solution of $MnO_2$ with constant magnetic stirring at 40 °C for 6 h. Afterwards, the 100 mL, 1 mM $KMnO_4$ aqueous solution was gradually poured in formerly stirred solution and was mixed under magnetic stirrer for further 30 min. A precipitate was collected at the end by centrifugation and rinsing consecutively with ethanol and DI water separately for 3 times with help of vacuum filtration. The powder obtained was then dried out in the vacuum oven (< 0.09 MPa) at 55 °C for 24 hours. The X-Ray diffraction (XRD) of the samples was done using the Bruker D8 Advance system. For obtaining elemental description, energy dispersive X-ray spectroscopy (EDX) was carried out using TESCON VEGA 3.

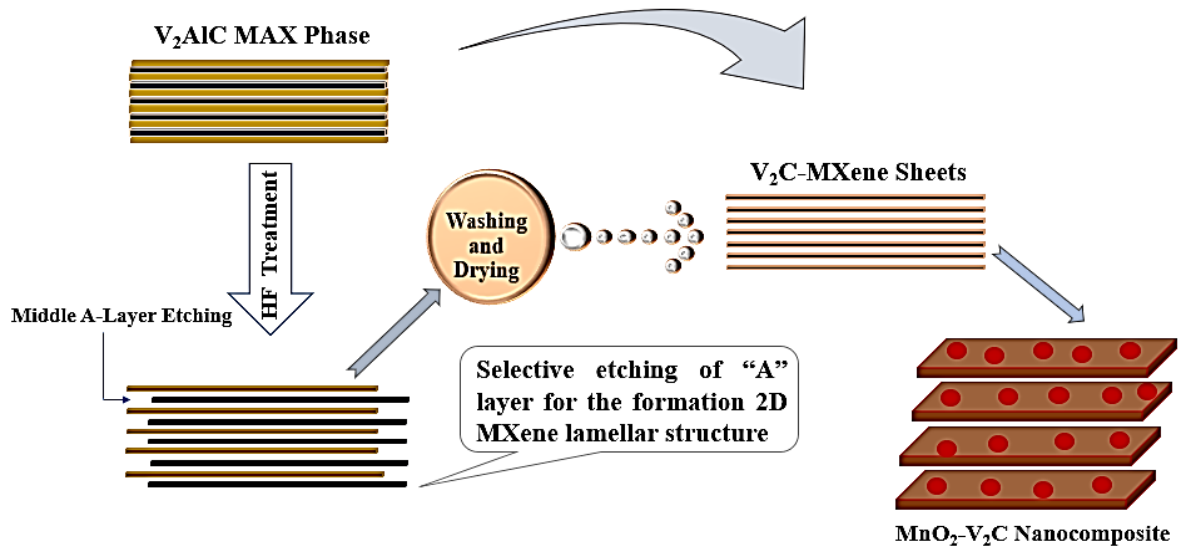

**Figure 1:** Schematic illustration of MnO$_2$-V$_2$C nanocomposite.

**Results and discussion**

The crystallographic information was obtained from XRD as shown in Fig 2a. The XRD pattern of V$_2$AlC, V$_2$C etched with 49% HF solution and 10% MnO$_2$-V$_2$C nanocomposite reveals a sharp peak at 2$\Theta$ = 13.28° and 41.09° of MAX precursor which shows its high crystallinity. The shifted peak is the consequence of an increased c-lattice parameter when MAX phase was etched and transformed into 2D MXene sheets. Moreover, the smaller peaks of MAX phase in MXene correspond to the presence of small unetched MXene [32–34].

XRD patterns of nanocomposite shows MnO$_2$ presence along with V$_2$CT$_x$. The broadened as well as shifted pattern of (002) diffraction peak to a lower angle suggested an increase in the interlayer spacing of the composite. The additional peaks in the MnO$_2$-V$_2$CT$_x$ nanocomposite at 2Θ of 35.5° and 39.6° are attributed to (112) and (101) polycrystalline planes of orthorhombic MnO$_2$ (JCPD 00-0300820) [35, 36]. Additionally, peak broadening has been observed which is due to the reduced crystallinity and presence of MnO$_2$ over V$_2$CT$_x$ sheets. Moreover, Figure 2c and 2d shows the micrographs of V$_2$C lamellar structure and MnO$_2$@V$_2$C MXene. The lamellar structure attained in Figure 2c has not been destroyed (Figure 2d) and persists even after the adsorption of MnO$_2$ on V$_2$C MXene. However, the interlayer spacing of V$_2$C sheets has evidently increased.

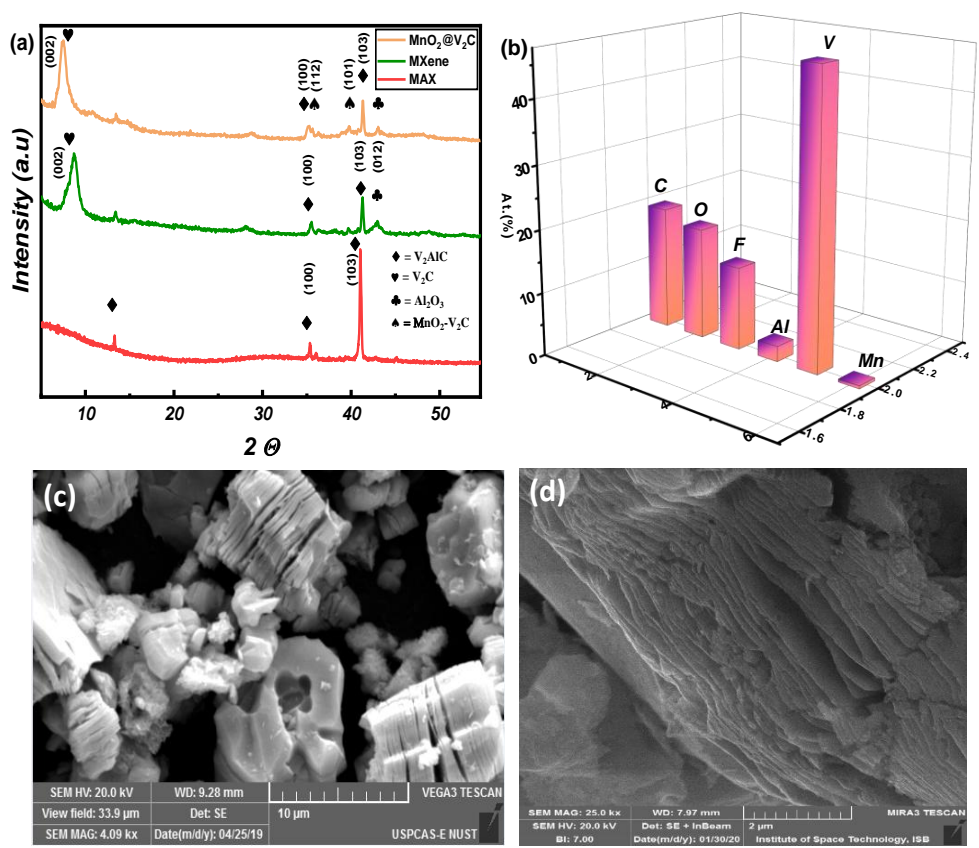

**Figure 2: a)** XRD of V$_2$AlC, prepared V$_2$CTx and MnO$_2$-V$_2$C nanocomposite **b)** EDS of MnO$_2$-V$_2$C nanocomposite, **c)** Micrograph of pristine V$_2$C **d)** Micrograph of MnO$_2$-V$_2$C nanocomposite.

Additionally, from the previous study, it is shown that pristine V$_2$C is generally metallic [24], characterized by a high density of states at the Fermi level and the metallic behavior is due to the delocalized electron states from the vanadium atoms, which contributes to high electrical conductivity. Furthermore, V$_2$C with surface terminated oxygen and Fluorine (V$_2$C-OF) can significantly alter its electronic structure. Oxygen atoms on the surface can introduce localized states that may reduce the metallic character, potentially creating a small bandgap or narrowing the density of states at the Fermi level. This change depends on the extent of functionalization. When Mn is adsorbed on the surface of V$_2$C-OF, it forms localized states associated with the Mn atoms and the oxygen functional groups. This adsorption modifies the

DOS near the Fermi level, depending on the strength of interaction between Mn and $V_2C$-OF. While Mn adsorption does not fully integrate into the lattice, it affects surface conductivity.

Moreover, the magnetic properties of Pristine $V_2C$ typically does not exhibit any intrinsic magnetism [23], as it lacks unpaired electrons that could generate a magnetic moment. The material is usually non-magnetic. However, $V_2C$-OF might exhibit slight magnetic effects, as oxygen functionalization can induce localized spin polarization in the V atoms. However, this magnetism tends to be weak or negligible, and $V_2C$-OF is often treated as either weakly magnetic or non-magnetic. Moreover, Mn adsorption induces localized magnetic moments on the surface, which leads to magnetic responses. Furthermore, the localized spin of Mn atoms results in ferromagnetic behaviour depending on Mn concentration and distribution. This is justified via detailed computational analysis described below.

**Crystal structure and Computational details**

The crystal structure of $V_2C$-OF and Mn adsorbed $V_2C$-OF is modeled by a supercell of slabs. For slab construction the bulk $V_2C$ structure was optimized to obtain the optimized lattice constants. This optimized structure was then used to construct the $V_2C$-OF surface slab. In Fig 3a, the carbon atom is sandwiched between vanadium layers. The O and F atomic layers were inserted to the system as surface terminations. A supercell of 2×2×1 was initially generated introducing vacuum of 11 Å to investigate the stability of Manganese (Mn) in slab using different positions of Mn as shown in Fig 3b-c. The internal geometry was further optimized with different Mn configuration. In internal geometry optimization the positions of the atoms were allowed to move in the direction of the force until the equilibrium has been attained. The doping and adsorption of Mn in $V_2C$-OF were observed to calculate formation energies and justify the Mn stability in the reported compound. Furthermore, the stable Mn-adsorbed $V_2C$-OF system was then studied in 4×4×1 supercell as shown in Fig 3d. The aim of the extended supercells in Figures 3(a-c) was to ensure that all cases were consistent with Figure 3(d).

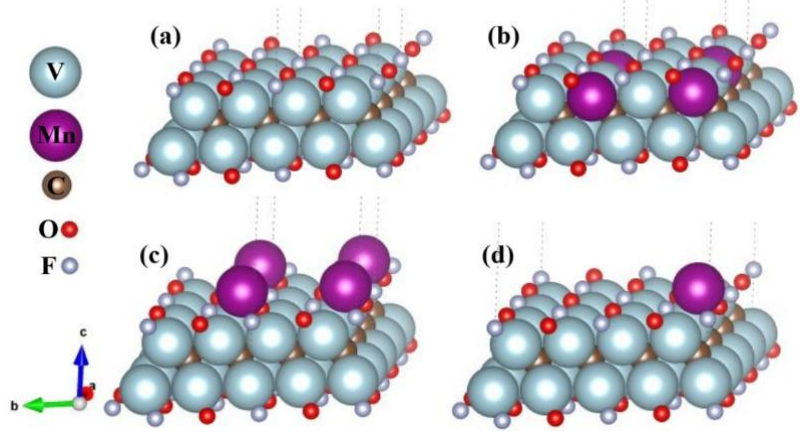

**Figure 3: a)** Structure of V$_2$C-OF **b)** Structure of Mn doped V$_2$C-OF in 2x2x1 supercell, **c)** Structure of Mn adsorbed V$_2$C-OF in 2x2x1 supercell **d)** Structure of Mn adsorbed V$_2$C-OF in 4x4x1 supercell.

The computational analysis was performed via ab-initio all-electron FLAPW method, as executed in the WIEN2k code [37]. The calculations were initiated using Pedrew-Burke-Ernzerhof (PBE) generalized gradient approximation (GGA) exchange-correlation functional for structure relaxation [38–40] . The single consistent field (SCF) calculation are converged using GGA+U with U= 3.0 eV for Mn-d state. In the interstitial regions, wave function was expanded in plane waves, and the plane wave cut-off chosen was $R_{MT}K_{max}$ = 7.0 where $R_{MT}$ represents the smallest radius of the atomic sphere and Kmax as the largest wave-vector magnitude. The $R_{MT}$ were taken as 1.86 a.u. for V-atoms, 1.55 a.u. for O-atoms, 1.68 a.u. for F-atoms, 1.63 a.u. for C-atoms and 1.80 a.u. for Mn-atom. For structure relaxation 2 k-points were used in irreducible brillouin zone with k-grid of 2×2×1. For energy convergence 54 k-points in IBZ with k grid of 6×6×3. Moreover, the forces relaxation criteria were kept at $10^{-4}$ Ryd and energy convergence criteria was fixed at $10^{-5}$ Ryd. V$_2$C-OF system is modelled by a supercell of slabs.

**Structure Stability**

In DFT, formation energy is a crucial parameter that significantly contributes to our understanding of the relative stability of various atomic substitutions within crystal structures and their implications in chemical reactions. The formation energy that can be calculated using the equations.

$$H_f^{V_{2-x}Mn_xC-OF} = E(V_{2-x}Mn_xC - OF) + E(V_2C - OF) + xE(V) - xE(Mn) \quad (1)$$

$$H_f^{Mn\ add-V_2C-OF} = E(Mn\ add - V_2C - OF) - E(V_2C - OF) - E(Mn) \quad (2)$$

The formation energy of Mn doped $V_2C$-OF structure, calculated by Equation (1), is 1.9014 eV/unit cell. For Mn-adsorbed $V_2C$-OF, it was calculated by Equation (2) and is -0.2388 eV/unit cell which clearly describes that Mn adsorbed $V_2C$-OF structure shows better stability than Mn doped $V_2C$-OF structure. Mn-adsorbed $V_2C$-OF system was then studied in 4×4×1 supercell. The schematic diagram of $V_2C$-OF with different configuration of Mn is shown in Figure 3(a-d). The $V_2C$-OF exhibits non-magnetic properties with the net magnetic moment (M) of about 0.14 µB per formula unit which is very small, and is consistent with the previous work [23, 41] whereas, M for Mn adsorbed $V_2C$-OF is found to be 4.48$\mu_B$. The increased M obtained in our calculations justifies the presence of strong ferromagnetism in Mn adsorbed $V_2C$-OF as compared to non-magnetic $V_2C$-OF.

Mn itself exhibits a ferromagnetic nature and readily forms a bond with oxygen. Thus, when Mn is adsorbed on the surface of $V_2C$-OF, manganese atom forms a bond with oxygen and fluorine. In Figure 3d, we discuss the attachment of functional groups with Mn atom. There is 1 O and 2 F's surrounding Mn-atoms. Due to the bond formation with the oxygen atom, the magnetic moment of Vanadium atom becomes small because the number of free electrons is reduced due to charge sharing. The interaction is however complex when Mn is adsorbed on its surface, but clearly, the huge difference between the magnetic moments is due to the shielding effect of the functional groups attached (oxygen and fluorine). As Oxygen has the valence shell configuration of $2p^4$ while Flourine has $2p^5$, so when there is an arrangement in which 2 F-atoms and 1 O-atom forms a bond with Vanadium atom and the adsorption of Mn-atom occurs at the surface, the density of states increases drastically as compared to other reasonable configurations [29]. Consequently, the overall magnetic moment of Mn-adsorbed $V_2C$-OF becomes higher than that of $V_2C$-OF. This increased magnetic moment is discussed in detail by the help of Figure 4.

Figure 4a reveals the electronic band gap of Mn-adsorbed $V_2C$-OF system. The zero electronic band gap shows that the electronic density of states is much higher and significant number of electrons are present in the conduction band. The density of states is obtained by Kohn-Sham eigen values calculation on a fine k-grid in the irreducible Brillioun Zone [42]. In Figure 4b, a peak is observed around -12eV to -11eV for both $V_2C$-OF and Mn adsorbed structures which usually occurs due to hybridization of s, p and d orbitals. Whereas in the region from -8eV to -2eV, the density of spin-up electrons is higher compared to the spin-down. At fermi-level, there is a little difference between spin up and spin down peaks which is then onwards persisting in the conduction band, i.e. for 0 to 1.6 eV for $V_2C$-OF system and 0 to 2 eV for Mn-adsorbed $V_2C$-OF.

The total density of states (TDOS) versus energy plot is as shown in Figure 4c. At low energy around -12eV to -11eV the peak is mainly formed by C-s along with small contribution of V-d state. The DOS peaks around -8 eV are mainly originated from induced magnetized F-p and O-p states that are attached as functional group with Mn atom. From -7 eV to -3 eV region bands are formed by nonmagnetic F- p and O-p state that are attached functional group to the Vanadium. Below -2 eV the V-d state is hybridized with O-p state. Also, from -2 eV to 8 eV the valance and conduction bands are formed by predominant V-d state. For a comprehensive analysis of the partial density of states, please refer to Figure I in the supplementary file.

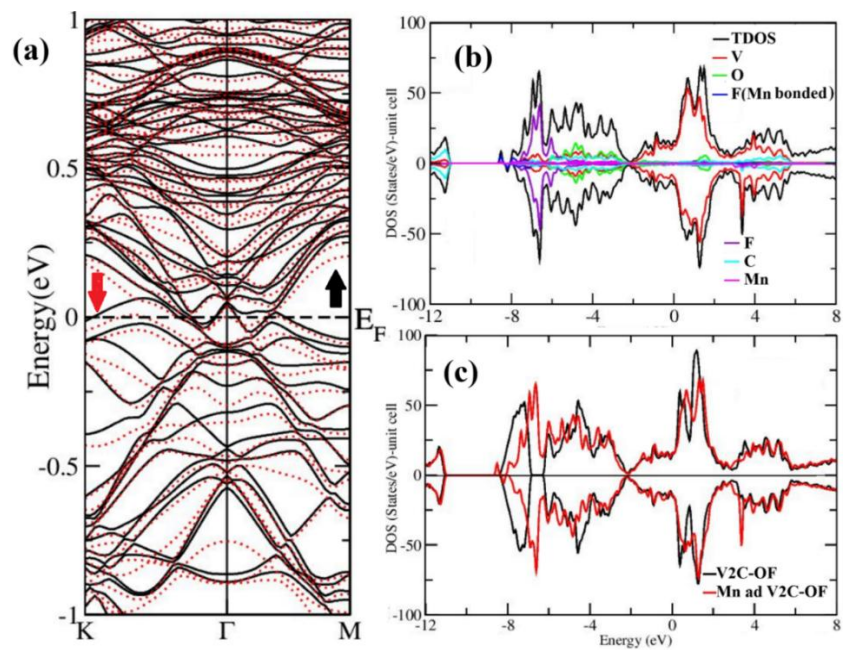

**Figure 4: a)** Electronic bandgap of Mn-adsorbed $V_2$C-OF **b)** Spin-polarized DOS Vs. Energy (eV) of $V_2$C-OF and Mn-adsorbed $V_2$C-OF system **c)** Total DOS Vs. energy (eV) of $V_2$C-OF and Mn-adsorbed $V_2$C-OF.

**Magnetic Analysis**

Magnetism in solids generally comes from localized electrons or delocalized electrons [42]. The interplay between these localized and delocalized electrons of in crystal structure defines the overall magnetic behavior of the solid. In this system, magnetism is induced by Mn-atoms in other elements specifically Fluorine atoms bonded with Mn. For more details of induce magnetization see table 1 in supplementary file. The F-p and O-p orbital are hybridized with Mn-d state results induce magnetization in F and O atoms. Also, the magnetic properties of Mn adsorbed $V_2$C originates from the d-orbitals of V-atoms. The V d-orbital electrons and localization of the d electrons by surface terminations (F, OH, H, or Cl) are also inducing magnetism in the compound [29].

In any case, magnetism comes particularly from exchange splitting subsequently resulting in a partial occupation of states, which differ between the spin-up (M↑) and spin-down (M↓) electrons. The corresponding magnetic moment $\mu_B$ is the difference between these occupation numbers (M = M↑ − M↓). In ferromagnetic metals, the order is collinear. Apart from this system, many elements for example, in Fe, Co, Ni shows collinear ordering and ferromagnetic nature [42]. In this work the total magnetic moment of the Mn ad $V_2C$-OF is found to be 4.48 $\mu_B$. Also, nonmagnetic-to-magnetic transition state relies upon applying a relatively small amount of strain. The 2D half-metallic $Ti_2C$ that is a ferromagnetic material and changes into a half-metal, a spin-gapless semiconductor, and then a metal under continuously applied biaxial strain. However, 2D $Ti_2N$ does not show any changes when a similar biaxial strain is applied to it [23].

To make the magnetic moment of V, O, F and C distinct, we show graph in Figure 5 up to 0.2$\mu_B$. We can observe the positioning of Vanadium atoms from 1 to 32 which shows that the magnetic moment has been induced in Vanadium atoms by the adsorption of Mn-atom. Consequently, the behavior of Mn-adsorbed $V_2C$-OF system overall is ferromagnetic. Also, the Oxygen atoms are labeled from 33 to 48 while Fluorine atoms are from 49 to 64. By the positioning of Oxygen and Fluorine atoms with Mn-atom, we can conclude that the magnetic moment of 2 Fluorine atoms and 1 Oxygen atom bonded with Mn-atom has induced in Fluorine and Oxygen atoms. The induced magnetization in Fluorine and Oxygen atoms is also shown in DOS figures (see Figure 4 and supplementary file)

The peaks in Figure 5 at atoms 41, 49 and 61 represent 1 Oxygen and 2 Fluorine bonded Mn atom. The $\mu_B$ of Mn-atom is about 3.99551$\mu_B$ however, the magnetic moments of other elements are comparatively very low. Therefore, to make the magnetic moment of other element visible we show Figure 5 on low scale. For clear description, Table 1 is given in the Supplementary File which directly shows that the magnetic moment of pristine $V_2C$-OF system is about 0.04034$\mu_B$. However, it is 4.48546$\mu_B$ for Mn adsorbed $V_2C$-OF system.

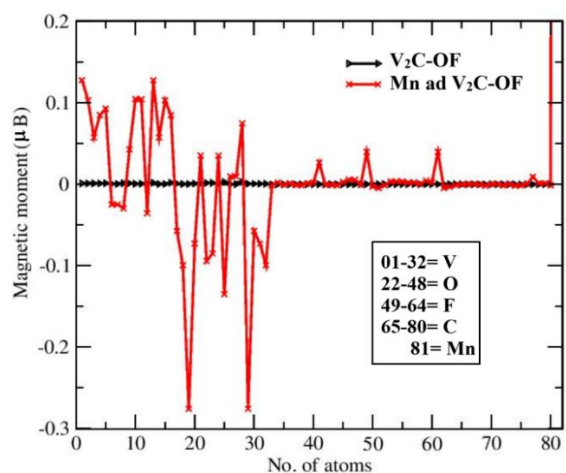

**Figure 5:** Computationally obtained Magnetic moments of $V_2C$-OF and Mn (ad) $V_2C$-OF systems.

**Conclusion**

The two-dimensional $V_2CT_x$ was synthesized from its parent compound MAX. This article reports the theoretical results on magnetic properties of pristine $V_2C$ and $MnO_2$ adsorbed $V_2C$ nanocomposite. XRD results showed that c-lattice parameter is increased from 13.01 Å to 22.6Å for $V_2AlC$, $V_2C$ and 27.2Å for $MnO_2$-$V_2C$ nanocomposite respectively. Clearly, signifying the adsorption-dominant properties. SEM and EDX signified the adsorption of $MnO_2$ in $V_2C$. In addition to it, the computational analysis revealed the strong ferromagnetic nature of $MnO_2$ adsorbed $V_2C$ while bare or $V_2C$-OF shows a non-magnetic nature. The current work will lead to the understanding of two-dimensional materials and investigates potential of MXenes for diverse applications in the field of 2D spintronics.

**Acknowledgment**

Authors are thankful to the Higher Education Commission (HEC) of Pakistan for provision of funding of research under the Project No.: 20-14784/NRPU/R&D/HEC/2021. Saleem Ayaz Khan acknowledges the support by the QM4ST project financed by the Ministry of Education of the Czech Republic grant no. CZ.02.01.01/00/22_008/0004572, co-funded by the European Regional Development Fund.

**Authors Contribution**

Mahjabeen Fatima analyzed computational data and wrote the manuscript, Saleem Ayaz Khan has extensively assisted in computational calculation and analysis, and Syed Rizwan conceived the research concept, assisted in manuscript writing and supervised the complete project.

**Funding Sources**

1. HEC: 20-14784/NRPU/R&D/HEC/2021
2. QM4ST CZ.02.01.01/00/22_008/0004572

**Conflict of interest**

There are no conflicts to declare.

## References

1 Fatima, M., Fatheema, J., Monir, N.B., Siddique, A.H., Khan, B., Islam, A., Akinwande, D., Rizwan, S. (2020) Nb-doped MXene with enhanced energy storage capacity and stability. *Frontiers in chemistry*, **8**, 168.

2 Hu, J., Xu, B., Ouyang, C., Zhang, Y., Yang, S.A. (2016) Investigations on Nb 2 C monolayer as promising anode material for Li or non-Li ion batteries from first-principles calculations. *RSC advances*, **6** (33), 27467–27474.

3 Iqbal, M.A., Ali, S.I., Amin, F., Tariq, A., Iqbal, M.Z., Rizwan, S. (2019) La- and Mn-Codoped Bismuth Ferrite/Ti3C2 MXene Composites for Efficient Photocatalytic Degradation of Congo Red Dye. *ACS Omega*, **4** (5), 8661–8668.

4 Sofo, J.O., Chaudhari, A.S., Barber, G.D. (2007) Graphane: A two-dimensional hydrocarbon. *Physical Review B*, **75** (15), 153401.

5 Cheng, Y., Wang, L., Li, Y., Song, Y., Zhang, Y. (2019) Etching and Exfoliation Properties of Cr2AlC into Cr2CO2 and the Electrocatalytic Performances of 2D Cr2CO2 MXene. *The Journal of Physical Chemistry C*, **123** (25), 15629–15636.

6 Das, S., Robinson, J.A., Dubey, M., Terrones, H., Terrones, M. (2015) Beyond graphene: progress in novel two-dimensional materials and van der Waals solids. *Annual review of materials research*, **45**, 1–27.

7 Pumera, M. and Sofer, Z. (2017) 2D monoelemental arsenene, antimonene, and bismuthene: beyond black phosphorus. *Advanced Materials*, **29** (21), 1605299.

8 Ren, X., Lian, P., Xie, D., Yang, Y., Mei, Y., Huang, X., Wang, Z., Yin, X. (2017) Properties, preparation and application of black phosphorus/phosphorene for energy storage: a review. *Journal of Materials Science*, **52** (17), 10364–10386.

9 Eklund, P., Beckers, M., Jansson, U., Högberg, H., Hultman, L. (2010) The Mn+ 1AXn phases: Materials science and thin-film processing. *Thin Solid Films*, **518** (8), 1851–1878.

10 Mannix, A.J., Zhou, X.-F., Kiraly, B., Wood, J.D., Alducin, D., Myers, B.D., Liu, X., Fisher, B.L., Santiago, U., Guest, J.R. (2015) Synthesis of borophenes: Anisotropic, two-dimensional boron polymorphs. *Science*, **350** (6267), 1513–1516.

11 Barsoum, M.W. (2013) *MAX phases: properties of machinable ternary carbides and nitrides*, John Wiley & Sons.

12 Barsoum, M.W. (2000) The MN+ 1AXN phases: A new class of solids: Thermodynamically stable nanolaminates. *Progress in solid state chemistry*, **28** (1-4), 201–281.

13 Huang, J.Y., Ding, F., Yakobson, B.I., Lu, P., Qi, L., Li, J. (2009) In situ observation of graphene sublimation and multi-layer edge reconstructions. *Proceedings of the National Academy of Sciences*, **106** (25), 10103–10108.

14 Barsoum, M.W. and Radovic, M. (2011) Elastic and mechanical properties of the MAX phases. *Annual review of materials research*, **41**, 195–227.

15  Wang, X.H. and Zhou, Y.C. (2010) Layered machinable and electrically conductive Ti2AlC and Ti3AlC2 ceramics: a review. *Journal of Materials Science & Technology*, **26** (5), 385–416.

16  Naguib, M., Kurtoglu, M., Presser, V., Lu, J., Niu, J., Heon, M., Hultman, L., Gogotsi, Y., Barsoum, M.W. (2011) Two-dimensional nanocrystals produced by exfoliation of Ti3AlC2. *Advanced Materials*, **23** (37), 4248–4253.

17  Naguib, M., Mashtalir, O., Carle, J., Presser, V., Lu, J., Hultman, L., Gogotsi, Y., Barsoum, M.W. (2012) Two-dimensional transition metal carbides. *ACS nano*, **6** (2), 1322–1331.

18  Coleman, J.N., Lotya, M., O'Neill, A., Bergin, S.D., King, P.J., Khan, U., Young, K., Gaucher, A., De, S., Smith, R.J. (2011) Two-dimensional nanosheets produced by liquid exfoliation of layered materials. *Science*, **331** (6017), 568–571.

19  25th Anniversary Article: MXenes: A New Family of Two-Dimensional Materials (2014), **26** (7).

20  Ghidiu, M., Naguib, M., Shi, C., Mashtalir, O., Pan, L.M., Zhang, B., Yang, J., Gogotsi, Y., Billinge, S.J.L., Barsoum, M.W. (2014) Synthesis and characterization of two-dimensional Nb 4 C 3 (MXene). *Chemical communications*, **50** (67), 9517–9520.

21  Harris, K.J., Bugnet, M., Naguib, M., Barsoum, M.W., Goward, G.R. (2015) Direct measurement of surface termination groups and their connectivity in the 2D MXene V2CT x using NMR spectroscopy. *The Journal of Physical Chemistry C*, **119** (24), 13713–13720.

22  Babar, Z.U.D., Fatheema, J., Arif, N., Anwar, M.S., Gul, S., Iqbal, M., Rizwan, S. (2020) Magnetic phase transition from paramagnetic in Nb 2 AlC-MAX to superconductivity-like diamagnetic in Nb 2 C-MXene: an experimental and computational analysis. *RSC advances*, **10** (43), 25669–25678.

23  Gao, G., Ding, G., Li, J., Yao, K., Wu, M., Qian, M. (2016) Monolayer MXenes: promising half-metals and spin gapless semiconductors. *Nanoscale*, **8** (16), 8986–8994.

24  Khazaei, M., Arai, M., Sasaki, T., Chung, C.-Y., Venkataramanan, N.S., Estili, M., Sakka, Y., Kawazoe, Y. (2013) Novel electronic and magnetic properties of two-dimensional transition metal carbides and nitrides. *Advanced Functional Materials*, **23** (17), 2185–2192.

25  Shein, I.R. and Ivanovskii, A.L. (2012) Planar nano-block structures Tin+ 1Al0. 5Cn and Tin+ 1Cn (n= 1, and 2) from MAX phases: Structural, electronic properties and relative stability from first principles calculations. *Superlattices and microstructures*, **52** (2), 147–157.

26  Enyashin, A.N. and Ivanovskii, A.L. (2013) Structural and electronic properties and stability of MX enes Ti2C and Ti3C2 functionalized by methoxy groups. *The Journal of Physical Chemistry C*, **117** (26), 13637–13643.

27  Lane, N.J., Barsoum, M.W., Rondinelli, J.M. (2013) Correlation effects and spin-orbit interactions in two-dimensional hexagonal 5d transition metal carbides, Tan+ 1Cn (n= 1, 2, 3). *EPL (Europhysics Letters)*, **101** (5), 57004.

28  Zhao, S., Kang, W., Xue, J. (2014) Manipulation of electronic and magnetic properties of M2C (M= Hf, Nb, Sc, Ta, Ti, V, Zr) monolayer by applying mechanical strains. *Applied Physics Letters*, **104** (13), 133106.

29  Fatheema, J., Khan, S.A., Arif, N., Iqbal, M., Ullah, H., Rizwan, S. (2020) Meissner to ferromagnetic phase transition in La-decorated functionalized Nb2C MXene: an experimental and computational analysis. *Nanotechnology*, **32** (8), 85711.

30  Iqbal, M., Fatheema, J., Noor, Q., Rani, M., Mumtaz, M., Zheng, R.-K., Khan, S.A., Rizwan, S. (2020) Co-existence of magnetic phases in two-dimensional MXene. *Materials Today Chemistry*, **16**, 100271.

31  Fatheema, J., Fatima, M., Monir, N.B., Khan, S.A., Rizwan, S. (2020) A comprehensive computational and experimental analysis of stable ferromagnetism in layered 2D Nb-doped Ti3C2 MXene. *Physica E: Low-dimensional Systems and Nanostructures*, **124**, 114253.

32  VahidMohammadi, A., Mojtabavi, M., Caffrey, N.M., Wanunu, M., Beidaghi, M. (2019) 2D MXenes: Assembling 2D MXenes into Highly Stable Pseudocapacitive Electrodes with High Power and Energy Densities (Adv. Mater. 8/2019). *Advanced Materials*, **31** (8), 1970057.

33  Naguib, M., Halim, J., Lu, J., Cook, K.M., Hultman, L., Gogotsi, Y., Barsoum, M.W. (2013) New two-dimensional niobium and vanadium carbides as promising materials for Li-ion batteries. *Journal of the American Chemical Society*, **135** (43), 15966–15969.

34  Dall'Agnese, Y., Taberna, P.-L., Gogotsi, Y., Simon, P. (2015) Two-dimensional vanadium carbide (MXene) as positive electrode for sodium-ion capacitors. *The journal of physical chemistry letters*, **6** (12), 2305–2309.

35  Feng, L., Xuan, Z., Zhao, H., Bai, Y., Guo, J., Su, C.-w., Chen, X. (2014) MnO 2 prepared by hydrothermal method and electrochemical performance as anode for lithium-ion battery. *Nanoscale research letters*, **9** (1), 1–8.

36  Fatima, M., Zahra, S.A., Khan, S.A., Akinwande, D., Minár, J. and Rizwan, S., 2021. Experimental and computational analysis of MnO2@ V2C-MXene for enhanced energy storage. *Nanomaterials*, *11*(7), p.1707.

37  Blaha, P., Schwarz, K., Madsen, G.K.H., Kvasnicka, D., Luitz, J. (2001) wien2k. *An augmented plane wave+ local orbitals program for calculating crystal properties*.

38  Schwarz, K., Blaha, P., Madsen, G.K.H. (2002) Electronic structure calculations of solids using the WIEN2k package for material sciences. *Computer physics communications*, **147** (1-2), 71–76.

39  Wimmer, E., Krakauer, H., Weinert, M., Freeman, A.J. (1981) Full-potential self-consistent linearized-augmented-plane-wave method for calculating the electronic structure of molecules and surfaces: O 2 molecule. *Physical Review B*, **24** (2), 864.

40  Perdew, J.P., Burke, K., Ernzerhof, M. (1996) Generalized gradient approximation made simple. *Physical review letters*, **77** (18), 3865.

41  Zhao, S., Kang, W., Xue, J. (2014) Role of strain and concentration on the Li adsorption and diffusion properties on Ti2C layer. *The Journal of Physical Chemistry C*, **118** (27), 14983–14990.

42  Blaha, P., Schwarz, K., Tran, F., Laskowski, R., Madsen, G.K.H., Marks, L.D. (2020) WIEN2k: An APW+ lo program for calculating the properties of solids. *The Journal of chemical physics*, **152** (7), 74101.